\documentclass[prl,aps,amsmath,twocolumn,showpacs,floatfix]{revtex4}
\usepackage{graphicx}
\begin{document}
%\draft
\title{Coulomb zero bias anomaly for fractal geometry and conductivity of
granular systems near the percolation threshold. }
\author{A.~S.~Ioselevich}
\affiliation{Landau Institute for Theoretical Physics RAS,  117940  Moscow, Russia,\\
Moscow Institute of Physics and Technology, Moscow 141700,
Russia.}
\date{\today}

\date{\today}

\begin{abstract}

A granular system
slightly below the percolation
threshold is a collection of finite metallic
clusters, characterized by wide spectrum of sizes, resistances,
and charging energies.  Electrons hop from cluster to clusters via
short insulating ``links'' of high resistance.
At low temperatures all clusters are Coulomb blockaded and the dc-conductivity $\sigma$ is exponentially suppressed. At lowest $T$ the leading transport mechanism is
variable range cotunneling via largest (critical) clusters,
leading to the modified Efros-Shklovsky law. At intermediate temperatures the principal
suppression of $\sigma$ originates from the Coulomb zero bias
anomaly occurring, when electron tunnels between adjacent large
clusters with large resistances. Such clusters are essentially
extended objects and their internal dynamics should be taken
into account. In this regime the $T$-dependence of $\sigma$ is
stretched exponential with a nontrivial index, expressed
through the indices of percolation theory. Due to the fractal
structure of large clusters the
 anomaly is strongly enhanced: it arises  not only in low dimensions, but also in
$d=3$ case.

\end{abstract}
\pacs{72.23.Hk, 73.22.-f, 72.80.Tm}

\maketitle

Granular materials play important role in modern technology and
material science (see \cite{book}). In the recent years
nanocomposite granular materials were invented, with
characteristic grain size $a$ on the scale of $1-10\; nm$
\cite{nanocomposites}. For systems with so small grains quantum
effects should be essential, in particular the Coulomb blockade
effect \cite{Devoret}. Therefore it is important to understand,
how the Coulomb blockade is manifested in realistic disordered
granular metal.

\begin{figure}
\includegraphics[width=0.9\columnwidth]{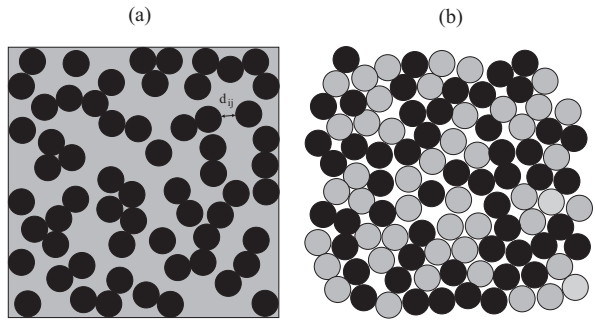}
\caption{Clusters of conducting grains in granular materials. (a)
-- Metal grains (shown black) in an insulating  matrix (shown
grey). Randomly distributed distances $d_{ij}$ between two
non-touching conducting grains determine corresponding tunnel
conductances $g_{ij}$. (b) -- A mixture of conducting  and
insulating grains. The ``links'' (one-grain insulating bridges)
between two conducting clusters are supposed to have the same
conductances $g_{ij}=g$, while all longer bridges have $g_{ij}\ll g$ and can be neglected.
 }
 \label{mixture}
\end{figure}

There are two large families of granular metallic materials. In
the systems of the first family
 conducting grains
are randomly embedded in an insulating matrix
(Fig.\ref{mixture}a), while the systems of the second family
 are mixtures of conducting and insulating grains
(Fig.\ref{mixture}b). The percolation
\cite{stauffer-aharony,bunde-havlin} is a general geometric
phenomenon, generic for  systems of both families. Some
conducting grains may touch each other \cite{contact},
establishing a  good contact (with dimensionless
conductance $G$), while the conductances $g_{ij}$ between grains
$i$ and $j$ which do not touch each other, are much smaller: $g_{ij}\ll G$.
If there is a
percolation via a network of touching each other conducting
grains, then electrons can travel throughout the system hopping from
grain to grain only via good contacts $G$. Otherwise hopping through some bad contacts $g$ is unavoidable.

If all conductances are small (both $G\ll 1$ and $g_{ij}\ll 1$),
then the Coulomb blockade effect exists at each metallic grain of
the system. It is characterized by charging energy on the scale $E_C^{(0)}\sim e^2/a$.
The mechanism
of the electronic transport at $T\ll E_C^{(0)}$ in
this case is  either direct intergrain hopping, described by the Arrhenius law
\begin{eqnarray}
\sigma\propto
  \exp\{-E_{\rm act}/T\}, \quad T_{\rm ES}\ll T\ll T_{\rm Arr},\label{activ1a66}
\end{eqnarray}
or {\it variable
range cotunneling} (VRC)\cite{ZhangShklovskii,FI05,bel-hopping2},
described by the modified Efros Shklovskii law
\cite{ES75}
\begin{eqnarray}
\sigma\propto \exp\left\{-\left(E_{\rm ES}/T\right)^{1/2}\right\}, \quad T\ll T_{\rm ES},
\label{es1d}
\end{eqnarray}
where
\begin{eqnarray}
T_{\rm Arr}\sim  E_C^{(0)},\quad T_{\rm ES}\sim  E_C^{(0)}/{\cal L}^*,\label{es1dtp}
\end{eqnarray}
\begin{eqnarray}
E_{\rm act}\sim  E_C^{(0)},\quad E_{\rm ES}\sim  {\cal L}(T)E_C^{(0)},\label{es1dtpp}
\end{eqnarray}

\begin{eqnarray}
{\cal L}^*\sim\ln(1/\overline{g}),
\quad {\cal L}(T)\approx {\cal L}^*+\ln\left[T_{\rm ES}^2/(T^2+T_{el}^2)\right].
\label{es1dtf}
\end{eqnarray}
Here $\overline{g}$ is a properly averaged intergrain conductance.
The temperature $T_{el}\sim [ E_C^{(0)}\delta]^{1/2}\ll T_{\rm ES}$
($\delta$ being the typical level spacing in a grain) corresponds
to the crossover from the inelastic cotunneling (at $T>T_{el}$) to
the elastic cotunneling (at $T<T_{el}$). The details can be found in \cite{FI05}.
Thus, ${\cal L}$ is a moderately large logarithmic factor and the Arrhenius
law can only be observed in a restricted intermediate temperature range.
 In the low-$G$ case  the presence or absence
of the percolation in the system is only relevant for the value
of $\overline{g}$, appearing in ${\cal L}^*$ as an argument of the $\log$-function,
and, therefore, it is only of a secondary importance.

In this paper we will study the large-$G$ case:
$
g_{ij}\ll 1$, but $G\gg 1$.
Here the mechanism of transport is  very sensitive to the percolation transition.
The resistance between two contacting metal grains is so low
that the charge easily spreads over  {\it clusters} of connected
metal grains (see Fig.\ref{mixture}), and it is clusters -- not
the individual grains -- that may either be Coulomb
blockaded, or not blockaded. If the fraction of the metal in the system $x$ is
larger than the percolation threshold $x_c$, then the Coulomb
blockade of finite clusters is not relevant, since the infinite
cluster of conducting grains exists (see
\cite{stauffer-aharony,bunde-havlin}), the current goes through
this infinite cluster, and electrons do not have to visit finite clusters whatsoever. It is not the case below
the percolation threshold, for $x<x_c$, where the
infinite cluster does not exist, and electrons have to  hop from one conducting cluster to
another due to tunneling through high-resistance insulating
bridges between them.

Close to the percolation threshold the distribution $N(n)$ of
numbers  of grains $n$ in a cluster has a long power-law tail:
$N(n)\sim n^{-\tau}$ at $1\ll n\ll n_{\rm cr}$. This tail is cut off
only at $n\sim n_{\rm cr}$, where
\begin{eqnarray}
 n_{\rm cr}\sim \xi^{d_f}\sim (x_c-x)^{-\nu d_f}\gg 1 \label{a1}
  \label{class2}
\end{eqnarray}
is the number of grains in a critical cluster, $\xi\sim
(x_c-x)^{-\nu}$ is its radius  (measured in the units of $a$),
$d_f$ is the fractal dimension of the infinite percolation cluster
(and of any large finite cluster with $n$ in the range $1\ll n\lesssim
n_{\rm cr}$ as well). The values of relevant critical exponents
are given in the table \ref{table1}. An estimate for typical
charging energy for a cluster consisting of $n$ grains
was found in \cite{ios-lyu}:
\begin{eqnarray}
E_C(n)\sim E_C^{(0)}n^{(\nu+s)/\nu d_f} \label{scale1fl}
\end{eqnarray}
 We
will see in what follows that the critical clusters with low
charging energy
\begin{eqnarray}
E_C^{\rm (cr)}\sim E_C(n_{\rm cr})\sim E_C^{(0)}(x_c-x)^{\nu+s}\ll
E_C^{(0)}
 \label{scale1f}
\end{eqnarray}
play the key role in low temperature transport at $x<x_c$.

\begin{table}[h]
\begin{tabular}{||c||c|c|c|c|c|c|c|c|c|c||}
\hline $d$ & $\nu$  & $d_f$ & $\tau$  & $\mu$ &
$s$  & $\Theta$ & $\tilde{\Theta}$ & $\varphi$\\
\hline \hline d=2 & 4/3  & 91/48 & 187/91   & 1.30 &   1.30  & 0.14 &  0.11 & 0.33\\
\hline d=3 & 0.875  & 2.524 & 2.32  & 2.14 &   0.74  & 0.38 &  0.08& 0.41\\
\hline
\end{tabular}
\caption{Numerical values of some critical
exponents}\label{table1}
\end{table}

It is natural to expect that the universal character of the
percolation transition, generic for most granular systems, should
lead to the essential universality  of the conductivity mechanism
for $x_c-x\ll 1$. One should, however, have in mind that besides
the topological disorder (that is responsible for the percolation
phenomena), there is yet another disorder: the randomness of
``bad'' conductances $g_{ij}$. Indeed, $g_{ij}\propto\exp(-2\kappa
d_{ij})$ where $d_{ij}$ is the separation between the two grains
and $\kappa$ is the tunneling decrement of the electronic wave
function in the insulator. If typical value of $\kappa d_{ij}\gg
1$, then the dispersion of $g$ is exponentially wide.

In  mixtures the distribution of $g$ is approximately discrete: the
thickness $d_{ij}$ of the insulating interval between two metal
clusters is, roughly, measured in the units of the diameter of the
insulating grain $d$. Then the tunneling conductances of shortest
one-grain insulating bridges (links) are $g_1\propto\exp(-2\kappa
d)$; the conductances of two-grain bridges are
$g_2\propto\exp(-4\kappa d)\ll g_1$, and so on. For such a system
one can simply ignore all  long insulating bridges and take into
account only the shortest -- the links, ascribing the same
conductance $g\equiv g_1$ to all of them. The conduction process
in the resulting system resembles the next nearest neighbor (NNN)
percolation (see \cite{ios-lyu} for detailed discussion of this
process).

For  metal grains, embedded in an insulating continuum,
 the distribution of $d_{ij}$ is essentially continuous.
 This fact introduces to the system   yet another
percolation-like physics, similar to that of the standard hopping
conductivity (see, e.g., \cite{EfrosShklovsky78}). In the present
paper we do not consider this facet of the problem explicitly, so
that, strictly speaking, the consideration below is directly
applicable only to mixtures. We expect, however, that the
principal universal features of conductivity, based on the
universal properties of the clusters distribution near the
threshold, will be present also for the systems with continuous
insulating matrix.

\begin{figure}
\includegraphics[width=0.5\columnwidth]{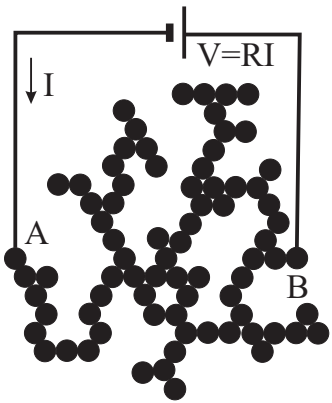}
\caption{Definition of the classic resistance $R$ across a
cluster. External voltage $V$ is applied  to  points $A$ and
$B$ at the periphery of the cluster, then the ratio of $V$ and the
current $I$ in the external circuit is the resistance $R$. }
 \label{clusters-conductance}
\end{figure}

The temperature dependence of conductivity of a mixture in the
intermediate range of temperatures
\begin{eqnarray}
E_C^{\rm (cr)},E_C^{\rm (m)}\ll T \ll E_C^{(0)}, \label{scale1i}
\end{eqnarray}
  was
already studied in the previous paper \cite{ios-lyu}. Here
\begin{eqnarray}
E_C^{\rm (m)}\sim E_C^{(0)}G^{-(s+\nu)/[\mu+(2-d)\nu]}\ll
E_C^{(0)} \label{scale1i1}
\end{eqnarray}
is the charging energy of a ``marginal cluster'', whose classic
resistance $R_m\sim 1$. The classic resistance across a fractal
cluster of $n$ grains (see Fig.\ref{clusters-conductance} and
Refs. \cite{bunde-havlin,ios-lyu}) is
\begin{eqnarray}
R(n)\sim G^{-1}n^{[\mu+\nu(2-d)]/\nu d_f},\label{cross-cond}
\end{eqnarray}
so that the number of grains in the marginal cluster is
\begin{eqnarray}
n_{m}\sim G^{\nu d_f/[\mu+(2-d)\nu]}\gg 1. \label{cross-scale}
\end{eqnarray}

The clusters with $n<n_{m}$ (and, therefore, with $R<1$) can
be treated as structureless {\it point-like} supergrains
characterized by the unique quantity -- the charging energy
$E_C(n)$. Under the condition (\ref{scale1i}) that is true for all
relevant clusters in the system and the conductivity is described
(see \cite{ios-lyu}) by the formulas
\begin{eqnarray}
 \sigma_{\rm ins}(x,T)\sim \sigma_{\rm ins}^{(0)}(x)
 [T/E_C^{(0)}]^{\Theta}, & \quad   T\ll T_{\rm cross}(x),
\label{freezing3ae}\\
 \sigma_{\rm cross}(T)\sim  \sigma_{\rm cross}^{(0)}
[T/E_C^{(0)}]^{\Theta'},  & \quad   T\gg T_{\rm cross}(x),
\label{freezing3aey}
\end{eqnarray}
where
\begin{eqnarray}
T_{\rm cross}(x) \sim E_C^{(0)}\left[(x_c-x)/\Delta_{\rm
cross}\right]^{(\mu+s)/\Theta}\label{fy1}
\end{eqnarray}
is the temperature of the crossover from the ``insulator-controlled'' conduction mode,
 where the resistivity is dominated by the insulating
links, to the ``critical crossover'' mode, where the voltage drops
occur  both on links and on the conducting
clusters. Note, that in the latter regime the conductivity
$\sigma_{\rm cross}$ does not  depend on $x$. The critical
exponents
\begin{eqnarray}
 \Theta=\frac{(d-2)\nu+s-1}{\nu +s},\quad \Theta'=\frac{\mu\Theta }{\mu+s}
\label{freezing3at}
\end{eqnarray}
are given in the table \ref{table1}.
At high temperatures $T\sim E_C^{(0)}$, when the Coulomb blockade effect
becomes irrelevant, the expressions
(\ref{freezing3ae},\ref{freezing3aey}) match with the known
results (see \cite{es76}) obtained in the absence of the Coulomb
effects:
\begin{eqnarray}
\sigma_{\rm ins}^{(0)}\sim g(x_c-x)^{-s},  \quad (\Delta_{\rm
cross}\ll x_c-x\ll 1),\label{class3}\\
 \sigma_{\rm cross}^{(0)}\sim g^{\mu/(\mu+s)}G^{s/(\mu+s)},  \quad (x_c-x\lesssim  \Delta_{\rm
cross}), \label{class3e}
\end{eqnarray}
where the width of the critical crossover domain of concentrations
is
\begin{eqnarray}
\Delta_{\rm cross}=(g/G)^{1/(\mu+s)} \label{class30}
\end{eqnarray}
The physics behind the results
(\ref{freezing3ae},\ref{freezing3aey}) is as follows: In the
temperature range \eqref{scale1i} it is possible to find
conducting paths consisting of {\it only large} metal clusters
connected by links, so that an electron never visits the
Coulomb-blockaded small clusters with $n<n_{\rm CB}(T)$ (and with charging energies $E_C(n)>T$). Provided $n_{\rm
CB}(T)<n_{\rm cr}$, the NNN-percolation in the system persists
despite the fact that all small clusters with $n<n_{\rm CB}(T)$
are not available for travelling electrons. This fact is by no
means trivial, since almost all conducting grains in the system belong to small clusters.
The power-law suppression factor
$(T/E_C^{(0)})^{\Theta}$ in the conductivity reflects just the
reduction of the number of available conducting paths.

In the present paper we will address the following questions:
\begin{enumerate}
\item What is changed if $T\ll E_C^{\rm (m)}$, so that the
relevant clusters are not point-like zero-dimensional, but
essentially extended objects with their internal dynamics? How one
should describe the Coulomb blockade phenomena in this case? \item
What is the conduction mechanism and what is the $T$-dependence of
conductivity at very low temperatures, when all clusters are
Coulomb blockaded?
\end{enumerate}

The low-temperature tunneling into a finite, but extended conductor
  is  suppressed  due to
 the process of charge spreading which transforms the initial
 point-like distribution of the tunneling charge into the smooth
 equipotential distribution. The corresponding suppression factor $\exp(-S_{\rm spr}(T))$
is nothing else,
 but the Coulomb zero-bias-anomaly (ZBA)
factor, which appears in the tunneling probability alongside with the usual Coulomb blockade factor
$\exp(-E_C/T)$. At low temperatures, when  the sample size $L$ is smaller
than the ``spreading length'' $L(T)$, the activation
 factor dominates, while at intermediate temperatures (for $L>L(T)$) the principal
 contribution  comes from the ZBA factor. Explicit expressions for $S_{\rm spr}(T)$
 (as well as for $L(T)$) are known for diffusive conductors of different
geometries: one-, two-, and three-dimensional (see
\cite{AA,Finkelstein,Nazarov,LevitovShytov,MishchenkoAndreevGlazman,EggerGogolin01,FI08});
the suppression is exponentially strong only in $1d$ case. In our
problem we deal with an unusual case: we have to find the
ZBA-factor for tunneling between two large clusters with {\it
fractal geometry}, where  the diffusion constant $D$, the conductivity $\tilde{\sigma}$, and the
dielectric constant $\varepsilon$ are scale-dependent. We will see
that for such a geometry the suppression is exponentially strong
in any dimension:
 in particular, in $2d$ and in $3d$.

 \begin{figure}
\includegraphics[width=0.9\columnwidth]{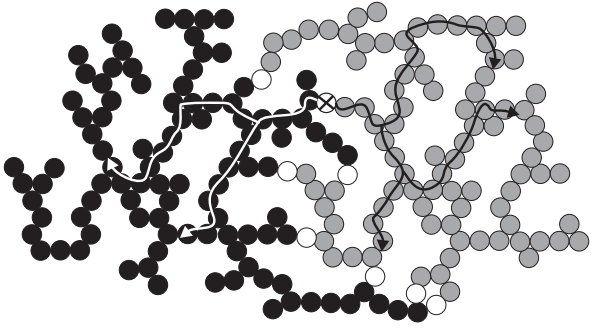}
\caption{The charge spreading process. After the tunneling of an electron  between two clusters (black and grey) through one of the links (marked by a cross) a diffusional spreading of charge occurs in both clusters.}
 \label{spreading}
\end{figure}

 Let us consider a tunneling of an electron through a link between
 two clusters with similar numbers of grains $n_1\sim n_2\sim n$.
  Immediately after the
intercluster hop the system finds itself under the barrier, with
large energy deficit $\sim E_C^{(0)}$, so that the charge density
has yet to spread over a large region in a tunneling manner,
before the system manages to get from under the barrier  (see Fig.\ref{spreading}). As a
result, the suppression factor has  the form $\exp\{-S(T)\}$ with
\begin{eqnarray}
S(T,n)=E_C(n)/T+S_{\rm spr}(T,n),\label{hyp4o}
\end{eqnarray}
where the  under-barrier action $S_{\rm spr}(T,n)$ can be
estimated by the semiclassical method. Different variants of such method were proposed in
\cite{Nazarov,LevitovShytov}, in this paper we will use the variant due to Levitov and Shytov \cite{LevitovShytov}.

Strictly speaking, the act of the electron hop between two
neighboring large clusters creates an electron-hole pair, and each
component of this pair then spreads over its own cluster. However, this
separation does not prevent the Coulomb interaction of both
components. The electron-hole interaction effectively leads to
partial screening of self-interaction in each cloud. If both
clouds develop in the same spatial domain, then this screening is
strong and leads to a dramatic suppression of the action  $S_{\rm
spr}$.
Exactly this situation arises,  when electron and hole clouds
proliferate in nearby parallel planes with similar properties. The
action in this case is parametrically smaller than that in the STM
case, where the hole is immediately evacuated \cite{LevitovShytov}.

In our problem we apparently have an intermediate case. The
neighboring large clusters to some extent interpenetrate, but by
no means they  coincide. The relative overlap is probably considerable, but definitely it is not close to a complete
coincidence.
Obviously, the effect of mutual screening is
important for our problem, but it only leads to an effective
suppression of self-interaction by a numerical factor, without
introducing any new scale or any new small parameter. Thus, we
conclude, that neglecting the mutual screening would overestimate
the under-barrier action $S_{\rm spr}$ only by a numerical factor
of order, say, two. Since we are anyway not able to determine the
numerical factor in the action, we will neglect the effect of
mutual screening in what follows.

We write the expression of the action $S_{\rm spr}$ in the
 case, when  the process is controlled by finite temperature (not by the external voltage), and
neglect the mutual  electron-hole screening:
\begin{eqnarray}
S_{\rm
spr}(T, n)\sim\sum_{k=0}^{\infty}\frac{2\pi T}{2\pi T(2k+1)+\tilde{D}_qq^2}\times\nonumber\\
\times\int_{1/L}^{1}\frac{d^d{\bf
q}}{(2\pi)^d} \frac{U_q}{2\pi T(2k+1)+\tilde{\sigma}_q q^2U_q+\tilde{D}_qq^2},
\label{levshe}
\end{eqnarray}
(see
\cite{FI08} for details).
Here $L$ is the size of an electrode, $\tilde{\sigma}$ and $\tilde{D}$ are its 
conductivity and diffusion constant, and $U$ is the screened Coulomb interaction.  The wave-vector $q$ is measured in the units of $1/a$. Now we extend the result \eqref{levshe} to the
fractal case with
\begin{eqnarray}
\tilde{\sigma}_q\sim Gq^{\mu/\nu}, \quad \tilde{D}_q\sim GE_{D}^{(0)}q^{d_f-d+\mu/\nu},\nonumber\\ U_q\sim
E_C^{(0)}q^{1-d+s/\nu},\quad L\sim n^{1/d_f}, \label{levsh1e}
\end{eqnarray}
(see \cite{bunde-havlin}). The  energy scale $E_{D}^{(0)}\sim E_C^{(0)}(ap_F)^{-2}\ll E_C^{(0)}$, therefore the diffusion terms $\tilde{D}_qq^2$ can be neglected in both denominators in \eqref{levshe}.
The sum over $k$ in
\eqref{levshed} is dominated by $k\sim 1$ and
the integrals over $q$  are dominated
by $q\sim L(T)^{-1}$, where
\begin{eqnarray}
L(T)\sim \left(GE_C^{(0)}/ T\right)^{\nu/[\mu+s+(3-d)\nu]}.
  \label{tun-depth}
\end{eqnarray}
The spreading factor is the dominant one,
if
$
1\ll L(T)\ll L$.
Introducing new variable
$Q=qL(T)(2k+1)^{-\nu/[\mu+s+(3-d)\nu]}$, and extending the
integration to the infinity, we get
\begin{eqnarray}
S_{\rm spr}(T, n)\sim
\left(E_C^{\rm (m)}/T\right)^{\varphi}, \label{levshed}
\end{eqnarray}
where
the new critical exponent $\varphi$ is expressed in terms of
the universal indices of the percolation theory:
\begin{eqnarray}
\varphi=\frac{\mu+(2-d)\nu}{\mu+s+(3-d)\nu}=\left\{
\begin{aligned}
0.33,\quad & (d=2),\\0.41,\quad & (d=3).
\end{aligned}
\right.\label{hyp4o1q}
\end{eqnarray}

\begin{figure}
\includegraphics[width=0.9\columnwidth]{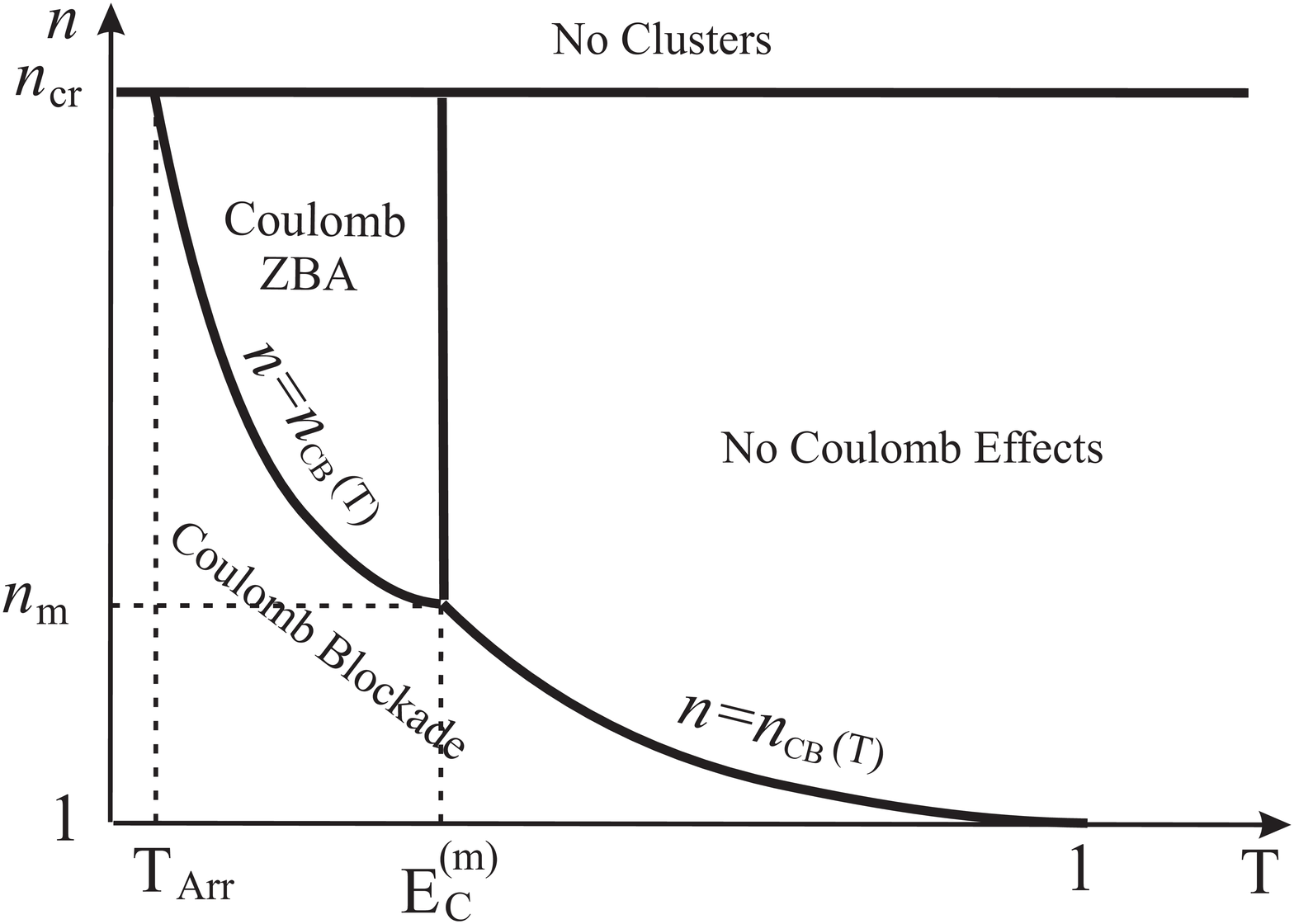}
\caption{Different regimes for tunneling between two clusters with
$n$ grains each on the $n-T$ plane. The most interesting case
$n_{\rm cr}>n_{\rm m}$ is shown; in the opposite case, for $n_{\rm
cr}<n_{\rm m}$, there are no clusters in the ``Coulomb ZBA''
domain. }
 \label{phase-diagram-clusters}
\end{figure}

The situation, when the action \eqref{hyp4o} is
dominated by the activation term, we call the ``Coulomb blockade'',
while the regime, dominated by the spreading term is called the ``Coulomb ZBA''.
The phase diagram for different
 regimes on the $n-T$ plane is shown in
Fig.\ref{phase-diagram-clusters}.
The crossover line is defined by
\begin{eqnarray}
n_{\rm CB}(T)\sim\left\{
\begin{aligned}
(T/E_C^{(0)})^{-\nu d_f/(\nu+s)},\quad & (T\gg E_C^{\rm
(m)}),\\(T/GE_C^{(0)})^{-\nu d_f/[\mu+s+(3-d)\nu]},\quad & (T\ll
E_C^{\rm (m)}).
\end{aligned}
\right.\nonumber
\end{eqnarray}
The crossover temperature $T_{\rm Arr}(x)$, below that all
clusters in the system are Coulomb blockaded, so that the
Arrhenius activated regime of conduction sets on, can be found
from the condition $n_{\rm cr}(x)=n_{\rm CB}(T_{\rm Arr})$:
\begin{eqnarray}
 \frac{T_{\rm Arr}(x)}{E_C^{(0)}}
\sim\left\{
\begin{aligned}
(x_c-x)^{s+\nu},\quad & (x_c-x\gg \Delta_{\rm
m}),\\G(x_c-x)^{\mu+s+(3-d)\nu},\quad & (x_c-x\ll \Delta_{\rm m}),
\end{aligned}
\right. \label{neu2att}
\end{eqnarray}
where
\begin{eqnarray}
\Delta_{\rm m}\equiv x_c-x_m\sim G^{-1/[\mu+(2-d)\nu]}
  \label{curt}
\end{eqnarray}
and $x_m$ is defined by the equation $n_{\rm cr}(x_m)=n_m$.

The result \eqref{levshed} is very different from the standard
Coulomb ZBA effect in plain (non-fractal) systems, where the
stretched exponential law
(with power
$1/2$) appears only in one-dimensional case
\cite{LevitovShytov,MishchenkoAndreevGlazman,EggerGogolin01,FI08}, while in 2-d
there is only a log-squared function of $T$ in the exponent and in
3-d the anomaly is weak. The reason for such a striking difference
is the nontrivial scale dependence \eqref{levsh1e} of both
dielectric screening and the conductivity
in the fractal system.

 In the range $T_{\rm Arr}(x)\ll T\ll E_C^{\rm (m)}$ the NNN-percolation
 via large non-blockaded clusters (with  $n_{\rm CB}(T)<n<n_{\rm cr}$), is still possible. However, the tunneling between these
clusters is strongly modified due to the Coulomb ZBA. As a consequence, the expressions for conductivity (\ref{freezing3ae},\ref{freezing3aey}) are  modified also:
The result
\eqref{freezing3ae} for the insulator controlled transport is changed to
\begin{eqnarray}
 \frac{\sigma_{\rm ins}(x,T)}{\sigma_{\rm ins}^{(0)}(x)}\sim
 \left(\frac{T}{E_C^{(0)}}\right)^{\tilde{\Theta}}B
 \exp\left\{-c\left(\frac{E_C^{\rm (m)}}{T}\right)^{\varphi}\right\},
\label{freezing3aer}
\end{eqnarray}
where $c$ is some unknown numerical constant, $B=B
 \left(T/E_C^{\rm (m)}\right)$ is unknown prefactor. Note that the
 expression on the right hand side of \eqref{freezing3aer} does not
 depend on $x$. The result
\eqref{freezing3aey} for the conductivity in the critical crossover range is changed to
\begin{eqnarray}
\frac{\sigma_{\rm cross}(T)}{\sigma_{\rm cross}^{(0)}}\sim
\left(\frac{T}{E_C^{(0)}}\right)^{\tilde{\Theta}'}
B^{\frac{\mu}{\mu+s}}\exp\left\{-\frac{\mu
c}{\mu+s}\left(\frac{E_C^{\rm (m)}}{T}\right)^{\varphi}\right\},
\label{freezing3aeyr}
\end{eqnarray}
and the expression for the crossover temperature \eqref{fy1}
is changed to
\begin{eqnarray}
T_{\rm cross}(x) \sim E_C^{\rm
(m)}\ln^{1/\varphi}\left[\Delta_{\rm
cross}^*/(x_c-x)\right],\label{fy1r}\\
 \Delta_{\rm cross}^*\sim \Delta_{\rm
 cross}G^{-\frac{(s+\nu)\tilde{\Theta}}{(\mu+s)[\mu+(2-d)\nu]}}.\label{fy1rd}
\end{eqnarray}
The modified critical exponents
\begin{eqnarray}
 \tilde{\Theta}=\frac{(d-2)\nu+s-1}{\mu +s+(3-d)\nu},
 \quad \Theta'=\frac{\mu\tilde{\Theta }}{\mu+s}.
\label{freezing3atr}
\end{eqnarray}

Now we turn to the question about the nature and the temperature
dependence of the conductivity at low temperatures $T<T_{\rm
Arr}$. Here an electron can not avoid Coulomb blockaded clusters,
since practically all clusters are blockaded.
At moderately low $T$ electrons travel through the ``critical
network of critical clusters''. This mode of transport is similar
to the standard nearest-neighbor-hopping regime in the hopping
conductivity (see \cite{EfrosShklovsky78}). The hopping goes via critical clusters, since, on
one hand, they have small charging energies $\sim E_C^{\rm (cr)}$,
and, on the other hand, they form an NNN-percolation network
(i.e., it is possible to travel by hopping from one critical cluster to another through direct links). In this
regime the conductivity obeys the Arrhenius law \eqref{activ1a66} with
$
 E_{\rm act}(x)\sim E_C^{\rm (cr)}$.

If the temperature is lowered further, then, in
the spirit of the Mott's variable range hopping
\cite{mott,EfrosShklovsky78} the variable range cotunneling
regime (VRC) \cite{ZhangShklovskii,FI05,bel-hopping2} sets on. This
regime is characterized by distant hops  between {\em resonant
clusters} with small charging energies. The hops are realized  as acts of multiple
cotunneling (elastic or inelastic) through chains of
nonresonant clusters, connecting the  resonant ones. Thus, VRC involves two sorts of
clusters:
\begin{itemize}
\item The terminal ones, where the real charged states can occur.
These clusters should be resonant. \item The
intermediate clusters, forming chains of virtual intermediate
states. These nonresonant clusters  should be as large as possible - to minimize the number
of intermediate states. Of course, they also should be
NNN-connected.
\end{itemize}

What is the nature of the resonant clusters? While the
characteristic scale of the charging energy for a cluster of size
$n$ is $E_C(n)$, its particular value for a given cluster is
random, due to random form of the clusters, and, most important,
due to random electrostatic potentials of the surrounding.
The electrostatic interaction of metallic clusters with charges
$Q_i$ (measured in the units of electronic charge $e$) is
$H_C\{Q\}=\frac{e^2}{2}\sum_{ij}\left[C^{-1}\right]_{ij}\tilde{Q}_i\tilde{Q}_j$, where $\tilde{Q}_i\equiv (Q_i-q_i)$ and
$C_{ij}$ is the matrix of capacitances.
 The so called
``background charges'' $-1/2<q_i<1/2$ are due to random potentials
of stray charges, trapped in the insulator; the set of $q_i$ is chosen so
that the ground state corresponds to $Q_i\equiv 0$. When one extra
electron (one extra hole) is put on the cluster $i$, the energy of
the system is changed by
\begin{eqnarray}
E_{Ci}^{(\pm)}=\frac{e^2}{2}\left\{\left[C^{-1}\right]_{ii}\mp
2\sum_{j}\left[C^{-1}\right]_{ij}q_j\right\}.
\label{cap1s}
\end{eqnarray}
The resonant clusters are characterized by anomalously small $E_{Ci}^{(+)}$ or $E_{Ci}^{(-)}$.
Counter-intuitively, the principal contribution to the low energy
density of states comes not from large resonant clusters, but from
small resonant clusters. To prove this, let us first neglect the long range Coulomb interaction (i.e., the
Coulomb gap effect) for a while, and consider the so called
``density of ground states''  $\nu_{GS}^{(0)}(E)$
(see \cite{ZhangShklovskii} for the detailed definition). It is
instructive to decompose it into the sum of contributions
$\nu_{GS}^{(0)}(E,n)$ of clusters with fixed $n$. Under the most natural
condition of strong charge disorder (when $q_i$ are homogeneously distributed in the interval
$-1/2<q_i<1/2$) each $\nu_{GS}^{(0)}(E,n)$ is
a structureless function with a single scale $E_C(n)$, so that
$\nu_{GS}^{(0)}(E,n)\sim 1/E_C(n)$ for $E\lesssim E_C(n)$ and
$\nu_{GS}^{(0)}(E,n)\approx 0$ for $E\gg E_C(n)$.
As a result, for the sum we obtain
\begin{eqnarray}
\nu_{GS}^{(0)}(E)= \sum_n
\nu_{GS}^{(0)}(E,n)N(n)\approx\nonumber\\\approx \sum_{n=1}^{n(E)}
n^{-\tau}/E_C(n)\approx \sum_n n^{-[\tau-(s+\nu)/\nu
d_f]}\sim 1. \label{parti1}
\end{eqnarray}
Here $n(E)$ is defined by $E_C(n)=E$. Since $\tau-(s+\nu)/\nu
d_f>1$, the sum  is indeed dominated by small clusters with $n\sim
1$. It is important to stress  that, although most clusters with
low charging energies are small resonant clusters, the latter do not
play any role in the conduction processes at $T>T_{\rm Arr}$.
Unlike large clusters, these  small resonant clusters are not
NNN-connected: they are separated from each other by many
nonresonant ones. That is why we did not take them into account in
the previous parts of this paper (as well as in \cite{ios-lyu}).

Taking into account the long range Coulomb interaction modifies
the result \eqref{parti1} and leads to the appearance of the soft
Coulomb gap
\cite{EfrosShklovsky78,ZhangShklovskii},
\begin{eqnarray}
\nu_{GS}(E)\sim |E|^{d-1}(\tilde{\epsilon}/e^2)^d,
\label{hype7edd}
\end{eqnarray}
but it can not alter the small-cluster nature of the majority of low energy
states.

The asymptotic law \eqref{hype7edd} is valid only for lowest
energies $|E|\ll E_C^{\rm (cr)}$, while in the range $E_C^{\rm
(cr)}\ll E\ll 1$ the density of states is strongly modified by the fractal character
of the system. However, the VRC regime is actual just in the
temperature range, where the  resonant clusters  have energies $E\ll E_C^{\rm (cr)}$, so that the
formula \eqref{hype7edd} is sufficient for our purposes.

The intermediate clusters involved in the acts of
multiple cotunneling are the critical ones.
In contrast with small resonant clusters, which are always
point-like, these  clusters can be either point-like (if
$n_{\rm cr}<n_{\rm m}$), or extended (if $n_{\rm
m}<n_{\rm cr}$). There is a considerable difference in the cotunneling
probability between the two cases.

Under the condition $n_{\rm cr}<n_{\rm m}$ all the critical
clusters at low temperature act as effective "supergrains" with
characteristic charging energy $E_C^{\rm (cr)}$. The average
number of links, connecting two adjacent critical clusters, is
$N_{\rm links}(\xi)\sim 1/(x_c-x)$ (see \cite{ios-lyu}). Therefore
the effective conductance between the two clusters is
\begin{eqnarray}
g_{\rm eff}\sim gN_{\rm links}(\xi)\sim g/(x_c-x).
\label{hype7edd1}
\end{eqnarray}
In this paper we  consider only the case $g_{\rm eff}\ll 1$. We also concentrate on the case
the inelastic cotunneling; a subtle question about the elastic
cotunneling and the $x$-dependence of $T_{el}(x)$ will be
discussed in a separate publication.
For point-like clusters the method of evaluation of the
probability of multiple cotunneling, proposed in \cite{FI05}, can
be applied directly. Introducing the characteristic
number $K$ of critical clusters, separating two resonant ones, and
the width $\Delta$ of the Mott strip, we obtain
\begin{eqnarray}
\sigma_{\rm ins}\propto g^K_{\rm eff}\left(\Delta/KE_C^{\rm
(cr)}\right)^{2K}\exp\left\{-\Delta/T\right\} , \label{hype7ed}
\end{eqnarray}
for $T_{el}(x)\ll T\ll T_{\rm Arr}(x)$. Then, having in mind the
relation $\Delta\cdot\nu_{GS}(\Delta) (\xi K)^d\sim 1$ between $\Delta$
and $K$, and optimizing the conductivity \eqref{hype7ed}, we
arrive at the formula \eqref{es1d} with $E_{\rm ES}\sim {\cal
L}(T)E_C^{\rm (cr)}$ and
\begin{eqnarray}
{\cal L}(T)={\cal L}^*+2\ln\left(T_{\rm ES}/T\right), \quad {\cal
L}^*\sim \ln \left(1/g_{\rm eff}\right). \label{es1}
\end{eqnarray}
Thus, for point-like critical clusters  $T_{\rm ES}\sim T_{\rm Arr}/{\cal L}^*$.

In the case $n_{\rm cr}<n_{\rm m}$ we have to deal with
cotunneling through extended objects with large resistance. Such
cotunneling was considered in \cite{FI08} for the case of long
diffusive wire;
the corresponding
exponential factor was found by means of modified
Levitov-Shytov semiclassical method. The approach of \cite{FI08}
can be applied also to the extended fractal clusters. Writing the
 action for the process, where an electron and
a hole are simultaneously injected into an extended cluster at its
opposite ends, at distance $r\sim \xi$ from each other, we obtain
\begin{eqnarray}
S_{\rm cotun}(T)\sim\sum_{k=0}^{\infty}\frac{2\pi T}{2\pi T(2k+1)+\tilde{D}_qq^2}\times\nonumber\\
\times\int_{1/\xi}^{1}\frac{d^d{\bf
q}}{(2\pi)^d} \frac{U_q\sin^2({\bf q}\cdot{\bf r}/2)}{2\pi
T(2k+1)+\tilde{\sigma}_q q^2U_q+\tilde{D}_qq^2}\sim\nonumber\\
\sim 2R(n_{\rm cr})\ln\left(T_{\rm Arr}/T\right)\label{levshe2}.
\end{eqnarray}
The expression \eqref{levshe2} differs from \eqref{levshe} only in
the factor $\sin^2({\bf q}\cdot{\bf r}/2)$.  In the ``Coulomb ZBA scenario''
 the integral over $q$ in \eqref{levshe} is cut off
at the intrinsic scale $q\sim L^{-1}(T)$, controlled by the
temperature and $x$-independent. On the contrary, in the VRC
regime $L(T)\gg\xi$, so that the integral is cut off at
$q\sim\xi^{-1}$.
As long as we are interested only in the inelastic cotunneling,
the diffusion  terms $\tilde{D}_qq^2$ again can be neglected in
both denominators in \eqref{levshe2}

Now, repeating the  arguments of the formula \eqref{hype7ed}
with account for the exponential suppression of the cotunneling
amplitude, we get again the result \eqref{es1d}, but with
$E_{\rm ES}\sim {\cal L}(T)E_C^{\rm (cr)}R(n_{\rm cr})$, $T_{\rm ES}\sim  T_{\rm Arr}/{\cal L}^*$ and ${\cal L}^*\sim 1/R(n_{\rm cr})\ln \left(1/g_{\rm
eff}\right)$.

Finally the results for both $n_{\rm cr}<n_{\rm m}$ and $n_{\rm cr}<n_{\rm m}$  can be unified in the following form: the crossover temperatures $T_{\rm Arr}$ and $T_{\rm ES}$, separaiting the domains of validity of the Arrenius law \eqref{activ1a66} and the Efros-Shklovskii law \eqref{es1d} are strongly reduced in the vicinity of percolation threshold,
\begin{eqnarray}
T_{\rm Arr}\sim  E_C^{\rm (cr)}/R_{\rm eff},\quad T_{\rm ES}\sim  T_{\rm Arr}/{\cal L}^*,\label{es1dtpy}
\end{eqnarray}
\begin{eqnarray}
{\cal L}^*=1+(1/R_{\rm eff})\ln \left(1/g_{\rm
eff}\right), \quad  R_{\rm
eff}\approx 1+R(n_{\rm cr}),\label{es2a}
\end{eqnarray}
\begin{eqnarray}
R(n_{\rm cr})\sim
G^{-1}(x_c-x)^{-1/[\mu+(2-d)\nu]}, \label{neu2a}
\end{eqnarray}
and $R_{\rm eff}$ has the meaning
of the effective dimensionless resistance  across
the critical cluster.

\begin{figure}
\includegraphics[width=1\columnwidth]{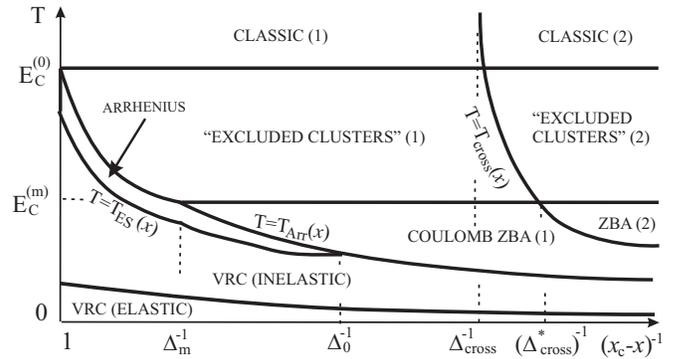}
\caption{Phase diagram of different regimes for the conductivity. ``Classic (1)'' -- classic
$T$-independent transport, controlled by the links (see Eq\eqref{class3});
``Classic (2)'' -- the same in
the critical crossover region (see Eq\eqref{class3e}); ``Excluded clusters (1)'' --
transport via large (non-blockaded) NNN-connected clusters,
the $T$-dependence is a power law (see Eq\eqref{freezing3ae});
``Excluded clusters (2)'' -- the same in the critical crossover region
(see Eq\eqref{freezing3aey}); ``Coulomb ZBA (1)'' -- transport is controlled
by the Coulomb zero-bias anomaly, the $T$-dependence is stretched
exponential (see Eq\eqref{freezing3aer}); ``ZBA (2)''
-- the same in the critical crossover region (see Eq\eqref{freezing3aeyr}); ``Arrhenius''
-- direct hopping between Coulomb blockaded critical clusters, the
$T$-dependence is simple activation (see Eq\eqref{activ1a66});
``VRC'' --  variable range cotunneling (elastic or inelastic) between
small resonant clusters via chains of critical non-resonant clusters,
the $T$-dependence is the modified Efros-Shklovskii law (see Eq\eqref{es1d}).}
 \label{phase-diagram-final}
\end{figure}

Note that the gap between  $T_{\rm Arr}$ and $T_{\rm ES}$ only exists,
if ${\cal L}^*\gg 1$. This condition is fulfilled for $x_c-x\gg \Delta_0$, where
\begin{eqnarray}
\Delta_0=\Delta_{\rm m}\left[\ln(1/g)\right]^{-[\mu+(2-d)\nu]},\label{es2bv}
\end{eqnarray}
while for $x_c-x<\Delta_0$ ${\cal L}^*\sim 1$ so that the gap
shrinks to zero and there is no room for the Arrhenius law (see Fig\ref{phase-diagram-final}).
The ``activation energies''  $E_{\rm act}$ and $E_{\rm ES}$ entering \eqref{activ1a66} and \eqref{es1d} are also reduced:
\begin{eqnarray}
E_{\rm act}\sim E_C^{\rm (cr)}, \quad E_{\rm ES}\sim {\cal L}(T)E_C^{\rm (cr)}R_{\rm eff},\label{es2b}
\end{eqnarray}
\begin{eqnarray}
{\cal L}(T)={\cal L}^*+2\ln\left(T_{\rm ES}/T\right),
\label{es1tt}
\end{eqnarray}

In conclusion, we have studied  the low-temperature behavior
of a granular system near the percolation threshold.
The peculiarities of this behavior stem from the strong dispersion and fractal properties of the conducting clusters.
 As in the system away from the percolation threshold, the transport mechanism at lowest temperatures
 is variable range cotunneling, and the corresponding temperature dependence of $\sigma$
 is the Efros-Shklovsky law. However, the parameters entering this law are dramatically
 renormalized in the vicinity of the threshold. In particular,  the onset of the VRC regime is shifted to lower temperatures. At higher temperatures the VRC
 is replaced by alternative phisical mechanisms. One of them -- the NNN-percolation with
 excluded small (Coulomb blockaded) clusters, was studied in \cite{ios-lyu}, the other --
 the Coulomb zero-bias anomaly scenario is first discussed in the present paper.
  The Coulomb
interaction suppresses the probability of tunneling between large metallic
clusters in the ``zero bias
anomaly'' manner. Due to the fractal structure of
 clusters the temperature dependence of the Coulomb
ZBA factor is described by stretched exponential law \eqref{freezing3aer} with nontrivial index
$\varphi$. 

The Author is indebted o M.V.Feigelman for numerous 
illuminating discussions and  comments which were crucial for this work. This research was
supported by RFBR grant \# 06-02-16533.


\begin{thebibliography}{99}
\bibitem{contact} The term ``touching'' does not necessarily mean
direct metal-metal contact. In some systems a thin insulating
(e.g., oxide) layer still separates two touching grains, but its
thickness  is much smaller than for
nontouching grains.
\end{thebibliography}
\end{document}